\documentclass[twocolumn,aps,prl,showpacs,amsmath,amssymb]{revtex4-1}
\usepackage{epsfig}
\usepackage{graphicx}
\usepackage{dcolumn}
\usepackage{bm}
\usepackage[colorlinks,citecolor=blue,linkcolor=red, urlcolor=blue,hyperindex]{hyperref}

\usepackage{dcolumn}

\usepackage{url}

\begin{document}
\title{Driving Driven Lattice Gases to Identify Their Universality Classes}

\author{Yahui Li}
 \thanks{These two authors contributed equally}
\author{Zhongda Zeng}%
 \thanks{These two authors contributed equally}

\author{Fan Zhong}%
 \email{ stszf@mail.sysu.edu.cn }
\affiliation{School of Physics, Sun Yat-Sen University, Guangzhou, 510275, P. R. China%
}%

\date{\today}

\begin{abstract}
    The critical behavior of driven lattice gas models has been studied for decades as a paradigm to explore nonequilibrium phase transitions and critical phenomena.
    However, there exists a long-standing controversy in the universality classes to which they belong. This is of paramount importance as it implies the question of whether or not a microscopic model and its mesoscopic field theory may possess different symmetries in nonequilibrium critical phenomena in contrast to their equilibrium counterparts.
    Here, we heat with finite rates two generic models of driven lattice gases across their respective nonequilibrium critical points into further nonequilibrium conditions. Employing the theory of finite-time scaling, we are able to unambiguously discriminate the universality classes between the two models. In particular, the infinitely driven lattice gas and the randomly driven lattice gas models belong to different universality classes. These results show that finite-time scaling is effective even in nonequilibrium phase transitions.
\end{abstract}

\maketitle

    Phase transitions and critical phenomena in nonequilibrium systems have attracted great concerns~\cite{Schmittmann,Marro,Hinrichsen,Odor,Chou,Henkel,zia2010}. Different from equilibrium systems, a systematic formalism for nonequilibrium systems is still lacking. It is thus indispensable to study simplified model systems. A genuine nonequilibrium system may have no equilibrium counterpart~\cite{Marro,Odor}. This includes surface growth models~\cite{EW,KPZ} and reaction-diffusion models that exhibit a phase transition into an absorbing state~\cite{Hinrichsen,Odor} such as models of epidemics~\cite{Mollison1977,Ligget1985}, catalysis~\cite{Ziff1986}, population~\cite{Albano1994}, and enzyme biology~\cite{Berry2003}. It can also be derived by modifying the parameters of an equilibrium system in such a way that a finite flux flows through it~\cite{Marro,Odor,Feng}. This latter category allows for making contact with known equilibrium systems for comparison.

    One of the possibly simplest models in this category is the driven lattice gas (DLG), which has been proposed to study the physics of fast ionic conductors under the influence of an electric field~\cite{katz1983,katz1984} and has become a prototype to study non-equilibrium statistical physics~\cite{Schmittmann,zia2010}. It starts from the Ising lattice gas model~\cite{ising1925}
     \begin{equation}\label{eq:ham}
        {\cal H} = -\sum_{ \langle {\bf i},{\bf j}\rangle} n_{\bf i} n_{\bf j},
    \end{equation}
    where the sum is over the nearest-neighbor pairs and $n_{\bf i}=1,0$ representing respectively the occupation or emptiness of site ${\bf i}$ in a $d$-dimensional space. ${\cal H}$ reduces to that of the standard Ising model after a mapping to spin variables via $\sigma_{\bf i} = 2 n_{\bf i} - 1$. The model evolves via nearest neighbor particle hole exchanges at a temperature $T$ so that the total particle number is conserved. Now, upon introducing a uniform external field ${\bf E}$ that biases the particle hopping and a periodic boundary condition along the field direction, a nonequilibrium steady state with a particle flux in the field direction is created. These conditions render the system topologically equivalent to an electric field looping around a torus, a situation which can be created by a time-dependent magnetic flux~\cite{Schmittmann}. Accordingly, the steady state cannot be described by a usual Gibbsian distribution with a Hamiltonian that includes the electric field itself and thus defies the general formalism of equilibrium statistics~\cite{katz1984}.

    The DLG on a half-filled lattice undergoes a continuous phase transition from a homogeneous disordered phase to an ordered phase with strips of alternate densities parallel to the field at a critical temperature $T_c$, which returns to the Ising critical point at $E=0$ and saturates for $E\rightarrow\infty$~\cite{katz1983,Schmittmann,Marro}, a case which is referred to as IDLG. Despite its seeming simplicity, the model defies general analytical solutions even in $d=1$. Yet, it exhibits many unexpected nonequilibrium behaviors such as negative responses, long-range correlations above $T_c$, and smooth interfaces below it~\cite{zia2010}. Nevertheless, the most peculiar aspect is that even the universality class of the transition is a subject of a long debate~\cite{Marro}, the aspect which leads to an important question as to whether collective behavior in nonequilibrium systems is governed by distinct principles.

    A usual method to describe the critical dynamics is to set up a coarse-grained Langevin equation and its corresponding mesoscopic field theory that respect the symmetry and conservation laws of the problem but ignore irrelevant microscopic details. For DLG, the standard Langevin equation for the conserved spin density $\phi({\bf r},t)$ is~\cite{JS1986,LC1986}
    \begin{equation}
              \partial_t\phi=\tau_\perp\nabla_\perp^2\phi-\nabla_\perp^4\phi+\lambda\nabla^2_\perp\phi^3
            +\tau_\parallel\nabla^2_\parallel\phi+{\cal E}\nabla_\parallel\phi^2+\zeta,
        \label{JSLC}
    \end{equation}
    where $\tau_\perp$ measures the distance to the critical point, $\nabla_\perp$ and $\nabla_\parallel$ are spatial gradients perpendicular and parallel, respectively, to the field, which is represented by ${\cal E}$ in a coarse-grained form, $\tau_\parallel$ and $\lambda$ are constants, and $\zeta({\bf r},t)$ is a conserved Gaussian noise. Renormalization-group (RG) analysis of the corresponding standard field theory (SFT) shows that the critical behavior is controlled by a nontrivial fixed point below an upper critical dimension $d_c=5$ determined by the most relevant nonlinearity term ${\cal E}$. Due to the Galileo invariance, the critical exponents can be exactly determined to all orders in the $\epsilon=d_c-d$ expansion for $d\leq5$~\cite{JS1986,LC1986}.

    However, early numerical results for $d=2$ found the order parameter critical exponent $\beta$ close to $1/3$, at odd with the SFT result of $1/2$~\cite{valles1986,valles1987,Marro1987,wang1989}. This led to a consideration that microscopic dynamics may be important for nonequilibrium systems~\cite{Garrido1995,munoz1994} and a new Langevin equation
    \begin{equation}
               \partial_t\phi= \tau_\perp\nabla_\perp^2\phi - \nabla_\perp^4\phi + \lambda\nabla^2_\perp\phi^3
            + \tau_\parallel\nabla^2_\parallel\phi + \zeta
    \label{RDLG}
    \end{equation}
    for IDLG was proposed~\cite{GSM1998antiJSLC,GSM2000antiJSLC,SMM2000antiJSLC}. This model coincides with a randomly-driven lattice gas (RDLG)~\cite{schmittmann1991RDLG,schmittmann1993RDLG}, where a parallel external field randomly changes its direction, and also a two-temperature Ising lattice gas~\cite{Garrido1990}. The critical behavior is well studied and controlled by the cubic term with a $d_c=3$ instead of $5$. The critical exponents have to be determined order by order due to the absence of the Galileo invariance and have been determined up to two-loop order in $\epsilon$. Moreover, the critical behavior of this nonequilibrium theory is described by the fixed-point equilibrium Hamiltonian of a uniaxial ferromagnet/ferroelectrics with nonlocal dipolar interactions~\cite{Schmittmann,schmittmann1993RDLG}. The most salient feature of Eq.~(\ref{RDLG}) is the absence of the field term and the ensuing particle current. As a consequence, particle-hole exchanges and space inversion, the symmetries of charge and parity (CP), respectively, are separately respected in Eq.~(\ref{RDLG}). This is in stark contrast to Eq.~(\ref{JSLC}) in which only their combined CP symmetry can retain the current direction.

    Although the disagreement of the early estimated $\beta$ with SFT was an incentive to the RDLG theory, anisotropic finite-size scaling (FSS)~\cite{fss,cardy1996FSS} with a fixed aspect ratio that accounts for the anisotropy resulted in opposite conclusions~\cite{leung1991,leung1992}. This was supported by subsequent studies~\cite{wang1996,JSLC2003FSS,caracciolo2004finite} albeit with questions~\cite{marro1996RDLG}. Nevertheless, an anisotropic FSS found that both the IDLG and RDLG models exhibit identical critical properties, indicating that it may be the strong anisotropy rather than the particle current that is the essential ingredient of the IDLG~\cite{RDLG2001FSS}. This raises an important question as to whether the symmetry of a microscopic model can be distinct from its continuous theory in nonequilibrium situations~\cite{schmittmann2000viability}. Therefore it is paramount to identify the universality of the IDLG model.

    Since anisotropic FSS results appeared disputed, one resorted to short-time critical dynamics in which critical behavior can be observed at early times and thus finite-size effects are expected to be negligible~\cite{Janssen,ZhengB,Albano}. However, two works along this line over a decade arrived at opposite conclusions~\cite{albano2002,daquila2012} with a debate~\cite{caracciolo2004comment,albano2004}. This controversy was resolved recently. It is shown that observables averaged along the field direction exhibit identical critical behavior in both the IDLG and RDLG models due to their effective Gaussian nature at short times~\cite{basu2017short,vo2017universal}. This indicates that the short-time method fails to discriminate the universality classes between the two models. Although a cumulant is found to show distinct long-time behaviors for the two models when their respective anisotropic exponents are given, the one for the IDLG demonstrates an FSS scaling without an explanation~\cite{basu2017short,vo2017universal,caracciolo2004finite}. Besides, other critical exponents remain untouched, especially $\beta$ that gives rise to the long time debate. These therefore necessitate other methods that can distinguish the two theories.

    A candidate is finite-time scaling (FTS) through manipulating dynamics~\cite{Zhong1,Zhong2}. This is to change the parameter of a system through its critical point at a finite rate $R$. Similar to the instantaneous change of parameters in DLG, such kinds of driving can also lead to a series of driven nonequilibrium critical phenomena such as negative susceptibility and competition of various regimes and their crossovers, as well as violation of fluctuation-dissipation theorem and hysteresis~\cite{Feng}. FTS is the temporal correspondence to FSS. The finite rate imposes a controllable external driving timescale of $\xi_R\sim R^{z/r}$ to the system, where $z$ and $r$ are respectively the dynamic critical exponent and the rate exponent associated with $R$~\cite{zhong2005dynamic,Zhong1}. This timescale characterizes
the time over which the field changes appreciably and plays a role similar to that of the system size $L$ plays in FSS near a critical point. As a result, FTS can bypass the notoriously critical slowing down of a divergent correlation time just as FSS can circumvent a divergent correlation length. It offers a detailed and improved understanding of the Kibble-Zurek scaling for the density of topological defects after cooling through a critical point~\cite{Kibble1,Kibble2,Zurek1,Zurek2,revqkz1,revqkz2}. FTS has been successfully applied to many systems to determine their critical properties~\cite{Zhong2,Yin3,Sandvik,Huang1,Liupre,Liuprl,Pelissetto,Xu,Xue,Gerster,Keesling}. It has also been combined with short-time critical dynamics to extend the Kibble-Zurek mechanism to beyond adiabaticity~\cite{Huang2}.
Here, we will employ FTS to drive the nonequilibrium transitions to further nonequilibrium to study their critical behavior. We find unambiguously that IDLG is well described by Eq.~(\ref{JSLC}) but not by Eq.~(\ref{RDLG}), which, instead, described the RDLG model reasonably. Therefore, the two models and their respective theories belong to different universality classes and their microscopic models and mesoscopic theories ought to share the same symmetry as in equilibrium.

We start with the FTS form of DLG. Consider an order parameter $M$. After a length rescaling of factor $b$, one assumes
    \begin{equation}
        M=b^{-\beta/\nu} M( t b^{-z}, \tau b^{1/\nu}, Rb^{r}, L_{\perp}^{-1}b,L_{\parallel}^{-1}b^{1+\Delta})\label{mb}
       \end{equation}
    for infinite drives, where $\Delta$ is the anisotropic exponent, $\tau=T-T_c$, $\nu$ is the critical exponent of the correlation length, and $L_{\perp}$ and $L_{\parallel}$ are the sizes perpendicular and parallel, respectively, to the field direction. We will always choose a time $t$ origin such that $\tau=Rt$. As a result, only two out of the three variables are independent. Moreover, because of the same relation and the rescalings of the variables involved in Eq.~(\ref{mb}), one obtains a scaling law $r = z + 1/\nu$~\cite{zhong2005dynamic}. In the absence of $R$, Eq.~(\ref{mb}) results from the RG theory~\cite{JS1986,LC1986}. In equilibrium transitions, a formal RG theory can be set up to account for $R$~\cite{Zhong06,Zhong1,Zhong2,Feng}. Here, we simply combine them as an ansatz and investigate its results. Setting $b=R^{-1/r}$ in Eq.~(\ref{mb}) leads to the characteristic timescale $\xi_R$ mentioned above. Removing further the redundant variable $t$, we arrive at the FTS form
    \begin{equation}
        M = R^{\beta/r\nu} f( \tau R^{-1/r\nu}, L_{\perp}^{-1}R^{-1/r},L_{\parallel}^{-1}R^{-(1+\Delta)/r}),
        \label{eq:Rop}
    \end{equation}
	where $f$ is a universal scaling function. We will employ Eq.~(\ref{eq:Rop}) to discriminate the universality classes between the IDLG and RDLG models at half filling and infinite drives in $d=2$.

\begin{table}
\caption{\label{tab:exponent} Critical exponents of the SFT~\cite{JS1986,LC1986} and the RDLG theory~\cite{schmittmann1993RDLG} in $d=2$.}
\begin{ruledtabular}
\begin{tabular}{ccccccc}
    &$\Delta$ &$\beta$  &$\nu$   & $\eta$ &$z$     &$r$  \\
\hline
SFT &$2$      &$1/2$    &$1/2$   &$0$     &$4$     &$6$\\
RDLG&$0.992$  & $0.315$ &$0.626$ &$0.016$ &$3.984$ &$5.581$\\
\end{tabular}
\end{ruledtabular}
\end{table}
To this end, we note that the critical exponents of the two theories have been well determined and those for $d=2$ are listed in Table~\ref{tab:exponent}. Accordingly, our task here is not to estimate them but rather to determine which theory describe the DLG model. This permits us to simplify Eq.~(\ref{eq:Rop}) by fixing two constants
${\cal C}_{\perp}=L_{\perp}^{-1}R^{-1/r}$ and ${\cal C}_{\parallel}=L_{\parallel}^{-1}R^{-(1+\Delta)/r}$. Consequently, Eq.~(\ref{eq:Rop}) is reduced to
\begin{equation}\label{eq:scalingfunction}
        M( \tau ) = R^{\beta/{r}\nu} f_M( \tau R^{-1/r\nu} ),
\end{equation}
where $f_M(X)=f(X,{\cal C}_{\perp},{\cal C}_{\parallel})$.
Similarly, the cumulant $g$ for an anisotropic system~\cite{cumulant,leung1991,caracciolo2004finite} follows a similar FTS form,
   \begin{equation}
        g=2-\langle M^4 \rangle / \langle M^2 \rangle^2 =f_G(\tau R^{-1/r\nu}),
        \label{eq:binder}
    \end{equation}
with another scaling function $f_G$. From Eq.~(\ref{eq:scalingfunction}), we can estimate $T_c$ from the cross point of $R^{-\beta/r\nu}M$ versus $T$ for a series of $R$ and thus $L_{\perp}$ and $L_{\parallel}$. It is somewhat unfortunate that the curves of $g$ for different $R$ coincide in low temperatures and do not cross and thus an unbiased estimate of $T_c$ is not available here. With the estimated $T_c$, curve collapses according to Eqs.~(\ref{eq:scalingfunction}) and (\ref{eq:binder}) can be invoked to test the results and identified the universality class.

To reduce the errors arising from noninteger values of the lattice sizes in fixing ${\cal C}_{\perp}$ and ${\cal C}_{\parallel}$, we first choose integers as candidates of $L_{\perp}$ and $L_{\parallel}$ that make the deviations from a constant aspect ratio ${\cal C}_{\perp}^{1+\Delta}/{\cal C}_{\parallel}$ as small as possible. With these possible lattice sizes, for a given ${\cal C}_{\perp}$, we then choose $R$, which also fixes ${\cal C}_{\parallel}$. We use Metropolis rates~\cite{MC} for hopping in the transverse directions. For an infinite drive, a particle always hops to the hole in the field direction, which changes randomly in each attempt for RDLG. Periodic boundary conditions are applied to both directions. As a narrow system is more likely to form an ordered phase with a single strip in its steady state~\cite{valles1986,2001OrderInitialState,albano2002}, we choose such a phase at $T=0$ as the initial configuration and heat it with the prescribed rate $R$. Each curve is averaged over $10~000$ samples.

    We choose a normalized anisotropic structure factor as the order parameter~\cite{leung1991,wang1996,daquila2012},
        \begin{equation}\label{eq:op}
            M(t) =\frac{ \sin(\pi / L_\perp) } {{2L_\parallel}} \left\langle \left|\sum_{j,k=1}^{L_\perp,L_\parallel} {\sigma_{j,k}(t)e^{i2\pi j/L_\perp}} \right|\right\rangle,
        \end{equation}
    which is normalized to unity for a completely ordered single strip, where the angle brackets denote ensemble averages. Because $2\beta/\nu=d-2+\Delta+\eta$~\cite{Schmittmann}, the right-hand side being the exponent for the structure factor, the definition gives rise to the correct exponent in Eq.~(\ref{mb}), where $\eta$ is the anomalous dimension.

    \begin{figure*}
		\centering
		\includegraphics[width=1\textwidth]{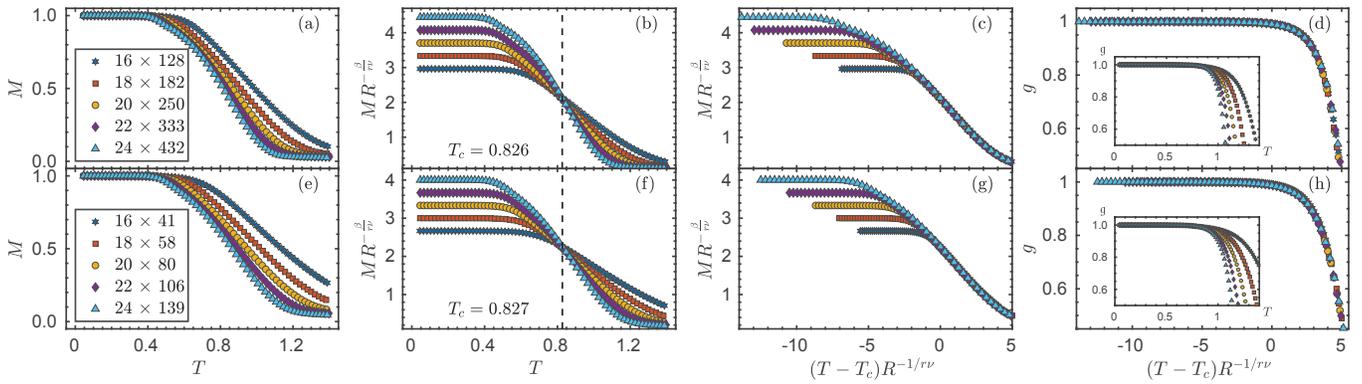}
		\caption{(Color online) $M$ and $g$ and their scalings of the IDLG model using the exponents of SFT for (a) to (d) ${\cal C}_{\perp}=5.4$ and ${\cal C}_{\parallel}=4.92$ and (e) to (f), ${\cal C}_{\perp}=6$ and ${\cal C}_{\parallel}=2.16$. The dash lines in (b) and (f) locate the critical temperature. The insets in (d) and (h) show the original $g$ curves. Each column shares identical abscissa and each row shares the same legend which lists $L_{\perp}\times L_{\parallel}$.}
		\label{fig.OP_j1st}
	\end{figure*}
    In Fig.~\ref{fig.OP_j1st}, we show the numerical results from IDLG with the exponents of SFT for two sets of ${\cal C}_{\perp}$ and ${\cal C}_{\parallel}$.
    It can be seen from Figs.~\ref{fig.OP_j1st}(a) and~\ref{fig.OP_j1st}(e) that the $M$ curves move to lower temperatures and become somehow steeper with smaller sweeping rates and larger lattice sizes. This trend is also exhibited by $g$ as shown in the insets of Figs.~\ref{fig.OP_j1st}(d) and~\ref{fig.OP_j1st}(h). These hystereses are a characteristic of the driven nonequilibrium critical phenomena~\cite{Feng} arising from the heating. Also $g$ equals $1$ in the ordered phase and tends to $0$ in the disordered phase, in agreement with previous studies~\cite{leung1991,leung1992,caracciolo2004finite,basu2017short,vo2017universal}. In Figs.~\ref{fig.OP_j1st}(b) and~\ref{fig.OP_j1st}(f), the curves of $R^{-\beta/r\nu}M$ versus $T$ cross as expected at $T_c\approx0.83$, which is estimated from the cross points that minimize the distances among the curves. With this $T_c$, both $M$ and $g$ collapse well after being rescaled as demonstrated in Figs.~\ref{fig.OP_j1st}(c),~\ref{fig.OP_j1st}(g),~\ref{fig.OP_j1st}(d) and~\ref{fig.OP_j1st}(h). No additional scaling factor is needed for $g$ as Eq.~(\ref{eq:binder}) dictates in contrast to the surprising results of FSS~\cite{caracciolo2004finite,basu2017short,vo2017universal}. We note that our $T_c$, which results in the best collapses, is slightly larger than $T_c\approx0.80$ reported in the previous works~\cite{katz1983,leung1991,caracciolo2004finite,daquila2012}. In fact, the latter also gives rise to a fairly good collapse. This is because the cross points in Figs.~\ref{fig.OP_j1st}(b) and~\ref{fig.OP_j1st}(f) have a minor rate dependence, which, in turn, stems probably from the possible logarithmic corrections~\cite{JS1986,LC1986,Schmittmann}.

    \begin{figure}[b]
		\centering
		\includegraphics[width=1\columnwidth]{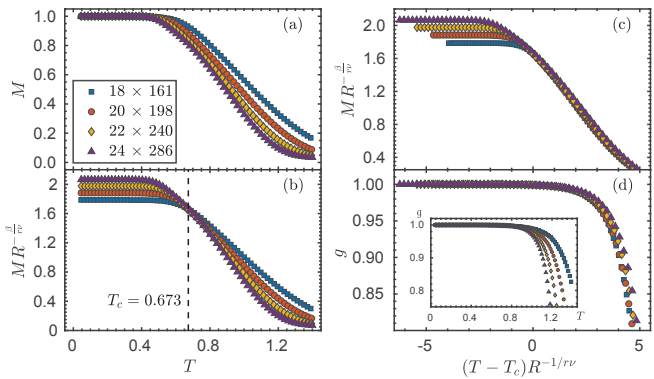}
		\caption{(Color online) $M$ and $g$ and their scalings of the IDLG model using the exponents of the RDLG theory instead of SFT for ${\cal C}_{\perp}=5.70$ and ${\cal C}_{\parallel}=16.3$. The dashed line in (b) marked the cross point, which is applied to (c) and (d) for curve collapses. The inset in (d) shows the original $g$ curves. All panels share the same legend.}
		\label{fig.OP_r1st}
	\end{figure}
	In Fig.~\ref{fig.OP_r1st}, we depict the numerical results from IDLG but using the exponents of the RDLG theory instead of SFT. As seen in Fig.~\ref{fig.OP_r1st}(b), the curves now nearly cross at $T_c=0.673$, which appears to collapse curves in Figs.~\ref{fig.OP_r1st}(c) and~\ref{fig.OP_r1st}(d). However, this apparent $T_c$ differs considerably from about $0.8$ in Fig.~\ref{fig.OP_j1st} and varies appreciably with ${\cal C}_{\perp}$ and ${\cal C}_{\parallel}$ opposite to the same $T_c$ found in Figs.~\ref{fig.OP_j1st}(b) and~\ref{fig.OP_j1st}(f). Also, this large difference cannot arise from the difference in lattice sizes. In addition, the quality of the collapses using such an apparent $T_c$ fluctuate, and those shown in Figs.~\ref{fig.OP_r1st}(c) and~\ref{fig.OP_r1st}(d) are poorer than their counterparts in Fig.~\ref{fig.OP_j1st}. Therefore, the transition of the IDLG model at $T_c\approx0.8$, which must be different from the one at $T_c\approx0.7$, if exists, cannot be described by the RDLG theory and does not belong to the universality class of RDLG.
	
	\begin{figure}[b]
		\centering
		\includegraphics[width=1\columnwidth]{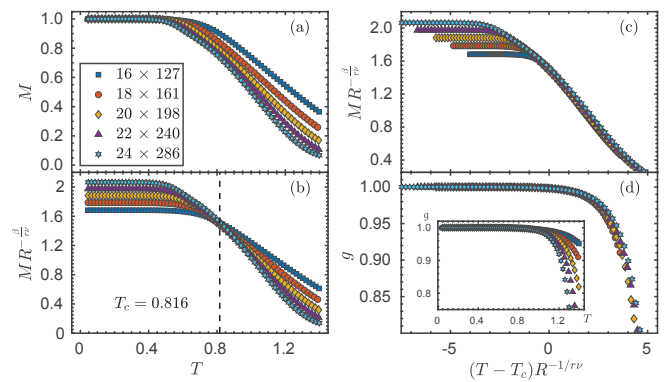}
		\caption{(Color online) $M$ and $g$ and their scalings of the RDLG model using the exponents of the RDLG theory for ${\cal C}_{\perp}=5.70$ and ${\cal C}_{\parallel}=16.3$. The dashed line in (b) marked $T_c$. The inset in (d) shows the original $g$ curves. All panels share the same legend.}
		\label{fig.OP_r2st}
	\end{figure}
    Finally, in Fig.~\ref{fig.OP_r2st}, we display the numerical results from RDLG using the exponents of its theory, Eq.~(\ref{RDLG}). From Fig.~\ref{fig.OP_r2st}(b), the rescaled curves of $M$ now cross approximately at $T_c\approx0.82$, with which both $M$ and $g$ collapse rather well as seen in Figs.~\ref{fig.OP_r2st}(c) and~\ref{fig.OP_r2st}(d). Our $T_c$ agrees with the previous value of $T_c\approx0.8$~\cite{RDLG2001FSS,albano2002,daquila2012,vo2017universal}, albeit slightly larger. In fact, the cross points in Fig.~\ref{fig.OP_r2st}(b) depend slightly on $R$. Also the quality of the collapses improves as the curves of the small lattice sizes are removed, though the corresponding cross points move to high temperatures. However, these may well originate from the fact that the critical exponents are directly computed from the two-loop series without resummations. Therefore, Fig.~\ref{fig.OP_r2st} shows that the RDLG exponents describe the RDLG model reasonably well.

    In conclusion, we have shown unambiguously that the critical behavior of the IDLG model is well described by SFT from Eq.~(\ref{JSLC}) rather than by Eq.~(\ref{RDLG}), which, instead, describes the critical behavior of the RDLG model reasonably. These, therefore, confirm at least for these driven diffusive systems that the symmetry of the microscopic models must be reflected in the mesoscopic field theories that are intended to correctly describe their collective behavior over large spatial and temporal scales. Our results show that FTS is effective even in these nonequilibrium transitions and thus driven nonequilibrium critical phenomena are also expected to exhibit in nonequilibrium critical phenomena.

\begin{acknowledgments}
This work was supported by the National Natural Science Foundation of China (Grant No. 11575297).
\end{acknowledgments}

\end{document}